\documentclass{article}
\usepackage{fleqn}
\usepackage{epsfig}
\usepackage{amsmath,amssymb}
\begin{document}

\begin{center}
{\bf\Large Numerical Simulation of Establishment of thermodynamic equilibrium in
cosmological model with an arbitrary acceleration}\\[12pt]
Yu.G. Ignat'ev\\
Kazan Federal University,\\ Kremlyovskaya str., 35,
Kazan 420008, Russia
\end{center}

{\bf keywords}: Early Universe, local thermodynamic equilibrium, relativistic kinetics,
scaling, cosmological acceleration, numerical simulation.

\begin{abstract}
Results of numerical simulation constructed before strict mathematical model of an establishment of thermodynamic equilibrium in originally nonequilibrium cosmological ultrarelativistic plasma for the Universe with any acceleration in the assumption of restoration of a scalling of interactions of elementary particles are presented at energies above a unitary limit. Limiting parametres of residual nonequilibrium distribution of nonequilibrium relic particles of ultrahigh energies are found.\end{abstract}

\section{Introduction}
\subsection{Mathematic Model}
In previous articles of the Author \cite{Ign1,Ign2} the strict mathematical model of restora\-tion
of thermody\-na\-mic equilibrium in originally nonequilibrium plasma in cosmological model with %
arbitrary acceleration has been constructed. The hypothesis about resto\-ra\-tion of a scalling
of interactions of elementary particles in the field of energies above a unitary limit that has %
led the Author to concept of a uniform asymptotic scattering cross-section (ACS), $\sigma_0(S)$, was the %
basic assump\-tion of the introduced model:
\begin{equation}\label{Yu15d}
\sigma_0(s)=\frac{8\pi\beta}{sL(s)},
\end{equation}
where $\beta\sim 1$,  $L(s)$ is a logarithmical factor. The analysis of results of numerical simulation has forced the Author to refuse the form of the logarithmic factor entered in late articles, and to return to the initial form of the logarithmic factor offered in article 1986 \cite{Yu_1986}:
\begin{equation}\label{yu15e}
L(s)=1+\ln^2\left(1+\frac{s_0}{s}\right)> 1,
\end{equation}
which is a monotone decreasing function of the kinematic invariant
$s$ --
$$\frac{d L}{ds}<0,$$
and $s_0=4$ is a squared total energy of two colliding Planck masses
so that on Planck energy scales:
\begin{equation}\label{Lambda(s0)}
L(s_0)\simeq1.
\end{equation}
This logarithmic factor differs from considered earlier on unit and is stabilized at energies above Planck. At energies below Planck addition of this unit does not change a little considerably mathematical model.

\subsection{Once again about an asymptotic scattering cross-section}
In the previous works arguments in favour of scalling behaviour of a scattering cross-section in the field {\it of energy above a unitary limit} have been resulted. Including it has been specified in coincidence of scatterings cross-section of some concrete four-partial reactions with values of the offered asymptotic scattering cross-section in corresponding areas of energies. However, in discussions with experts in the quantum theory of a field their unacceptance of this statement is often found out. According to the Author this unacceptance is caused, first of all, by rather limited range of energy in which are spent calculation of concrete scatterings cross-section, and, in-second turn, ``a sight from below'' in sense of energy below a unitary limit on quantum procedures of calculation of scatterings cross-section. For elimination of it of "small-scale misunder\-stan\-ding" the Author results in this article comparison of values of an asymptotic scattering cross-section (\ref{Yu15d}) with known quantum four-partial processes in gra\-phic formate (Fig. \ref{uacs_fig}).

\vskip 12pt
\centerline{\includegraphics[width=8.5cm]{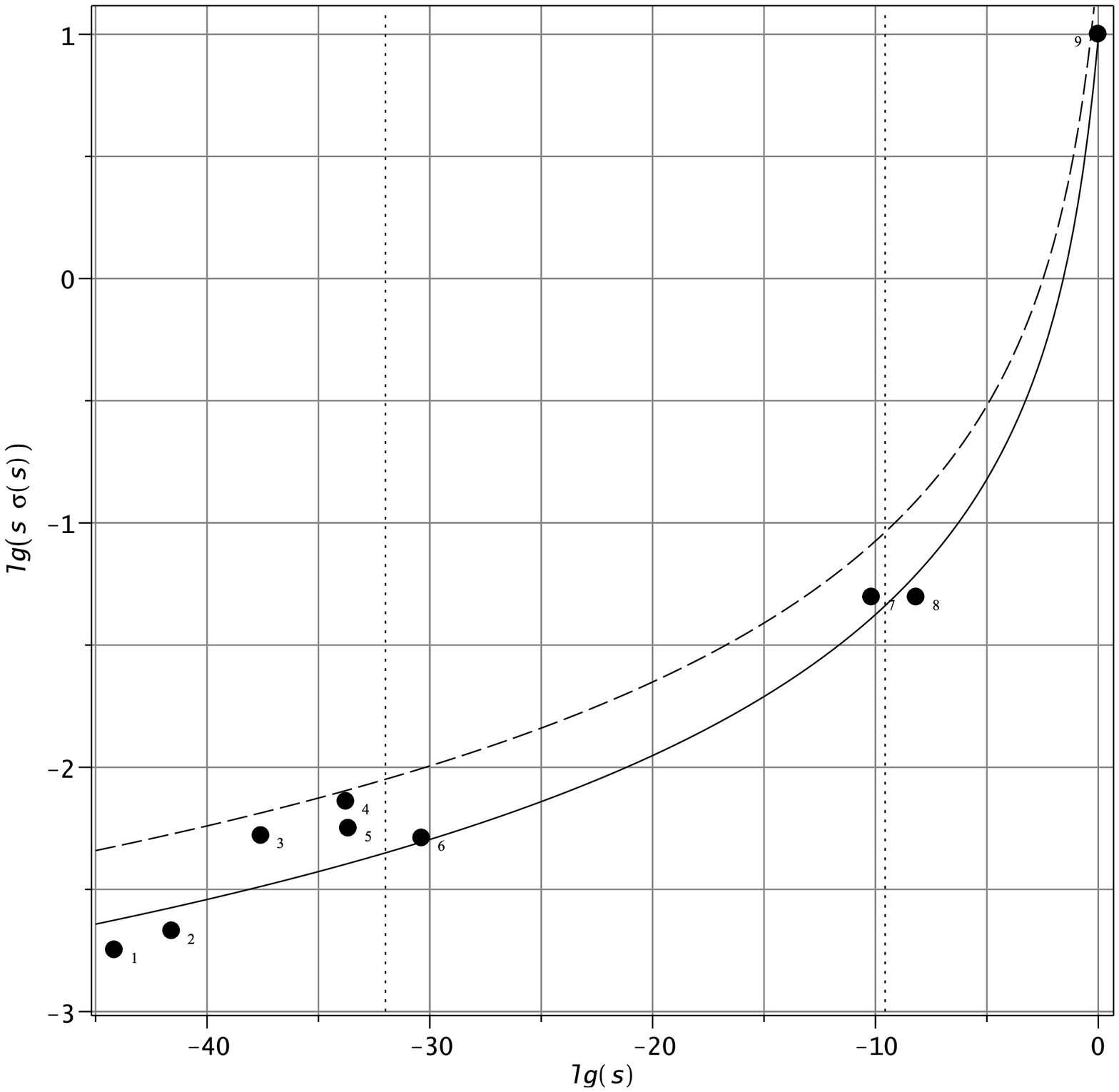}}
\refstepcounter{figure}{Figure \thefigure.}\label{uacs_fig}
\hskip 12pt {\small Comparison of the universal cross-section of
scattering (\ref{Yu15d}) at factor $\beta=1$ with the well-known
cross-sections of fundamental processes -- bold line. Dotted line
corresponds to the graph of universal cross-section of scattering at
factor  $\beta=2$. On the abscissa axis are laid values of the
common logarithm of the first kinematic invariant $\lg s$ in Planck
units; along the ordinate axis are laid values of the common
logarithm of the dimensionless invariant, $\lg s\sigma(s)$.  1 --
Thompson scattering, 2 -- Compton scattering on electrons at
$E_{SCM}=10$ Mev, 3 -- Compton scattering of electrons at
$E_{SCM}=1$ Gev, 4 -- electroweak interaction with participation of:
$W$ - bosons, 5 -- with participation of $Z$ - bosons, 6 -- with
participation of H-bosons at energy of the order of  7 Tev
($\sigma\sim 10$fb); 7 -- $SU(5)$ - interaction at mass of the
superheavy X-bosons $10^{15}$ Gev, 8 -- $10^{14}$ Gev; 9 --
gravitational interaction on Planck scales. Vertical dotte lines
correspond to energy values of the unitary limit for $SU(2)\times
SU(1)$ - interac\-tions, $E_u=617$ Gev, and SU(5) - interac\-tions,
$E_u\sim10^{14}$ Gev.} \vskip 10pt

\section{The numerical model of LTE restoration in the accelerating Universe}
\subsection{The model of the initial non-equilibrium distribution}
Thus, as was mentioned in papers \cite{Ign1,Ign2}, a mathema\-tical model of LTE
restoration process in cosmologi\-cal plasma is reduced to two
parametric equations
\begin{equation}\label{IV.57}
\int\limits_0^Z \frac{du}{\sqrt{1-(1-\sigma_0)\Phi(u)}}=\tau.
\end{equation}
\begin{equation}\label{IV.67}
y=[1-(1-\sigma_0)\Phi(Z)]^{1/4},
\end{equation}
which {\it at
given function} $\Phi(Z)$ define relations of form:
\begin{eqnarray}\label{tau_Z}
\tau=\tau(Z);\\
y=y(Z),
\end{eqnarray}
solving which we can determine function $y(\tau)$ and thereby
formally solve stated problem completely. Thus, the final solution
of the task is found in quadratures specifying the initial
distribution of non-equilibrium particles $\Delta f_0(p)$ and
following defi\-ni\-tion of the integral function $\Phi(Z)$:
\begin{equation}\label{IV.52}
\Phi(Z)\equiv {\displaystyle \frac{\sum\limits_a (2S+1)
\int\limits_0^\infty d\rho\rho^3\Delta f^0_a(\rho) {\rm
e}^{\displaystyle - \frac{Z(\tau)}{\rho}}}%
{\sum\limits_a (2S+1) \int\limits_0^\infty d\rho\rho^3\Delta
f^0_a(\rho) }}.
\end{equation}
Let us note that formally parametric equations
(\ref{IV.57}) and (\ref{IV.67}), as well as function's $\Phi(Z)$
definition, do not differ from the similar, obtained earlier by the
Author in articles \cite{Yu_1986}, \cite{LTE2}. The main new
statement is brought by the acceleration of the Universe and
consists in the relation $\tau(t)$:
\begin{equation}\label{IV.42}
\tau=2\int\limits_0^t \frac{\xi}{a}\ dt\ ,
\end{equation}

In order to construct a numerical model let us consider the initial
distribution of white noise type:
\begin{equation}\label{df0}
\Delta f_0(\rho)=\frac{A}{\rho^3}\chi(\rho_0-\rho),
\end{equation}
where $A$ is a normalization constan, $\rho_0>1$ is a dimensionless
parameter, $\chi(x)$ is a Heaviside step function, so that the
conformal energy density with respect to this distri\-bu\-tion is equal
to:
\begin{equation}\label{e_0}
\tilde{\varepsilon}^0_{ne}=\frac{\langle
\tilde{p}\rangle_0^4 A\rho_0}{32\pi^5}.
\end{equation}
Calculating function $\Phi(Z)$ relative to distribution (\ref{df0}),
we find:
\begin{equation}\label{Phi_Z}
\Phi(Z)=e^{-x}-x{\rm Ei}(x);\quad x\equiv \frac{Z}{\rho_0},
\end{equation}
where ${\rm Ei}(x)$ is an integral exponential function
\begin{equation}\label{Ei}
{\rm Ei}(x)=\int\limits_{-1}^\infty \frac{e^{-tx}}{t}dt.\nonumber
\end{equation}
\subsection{The results of numerical integration}
Thus, the problem is reduced to the numerical integration of the
system of equations (\ref{IV.42}), (\ref{IV.57}), (\ref{IV.67}).
Below the certain integration results are represented. Further
according to (79) \cite{Ign1}
\begin{eqnarray}\label{V.30}
\varepsilon_s={\rm Const}=\frac{3\Lambda^2}{8\pi};\\%
\label{V.18}%
\varepsilon_p a^4 \equiv \tilde{\varepsilon}_p={\rm Const}\simeq
\frac{3}{32\pi}.
\end{eqnarray}
and (83) \cite{Ign1}
\begin{equation}\label{V.21}
a(t)=\frac{1}{\Lambda}\sqrt{\frac{3}{32\pi}{\rm sh} \frac{t}{2\Lambda}}
\end{equation}
it may be convenient to
introduce a {\it time cosmological constant}
\begin{equation}\label{t_0}
t_0\equiv 4\Lambda.
\end{equation}
In article \cite{Ign5} the Author's software package in system of computer mathematics Maple v.15
numerical simulation presented above mathematical model of restoration of thermodynamic
equilibrium in the Universe containing transition to a stage of acceleration is described.
More low we will describe results of numerical modelling in more details and we will carry
out their analysis.

\begin{figure}[h]
 \centerline{\includegraphics[width=8cm]{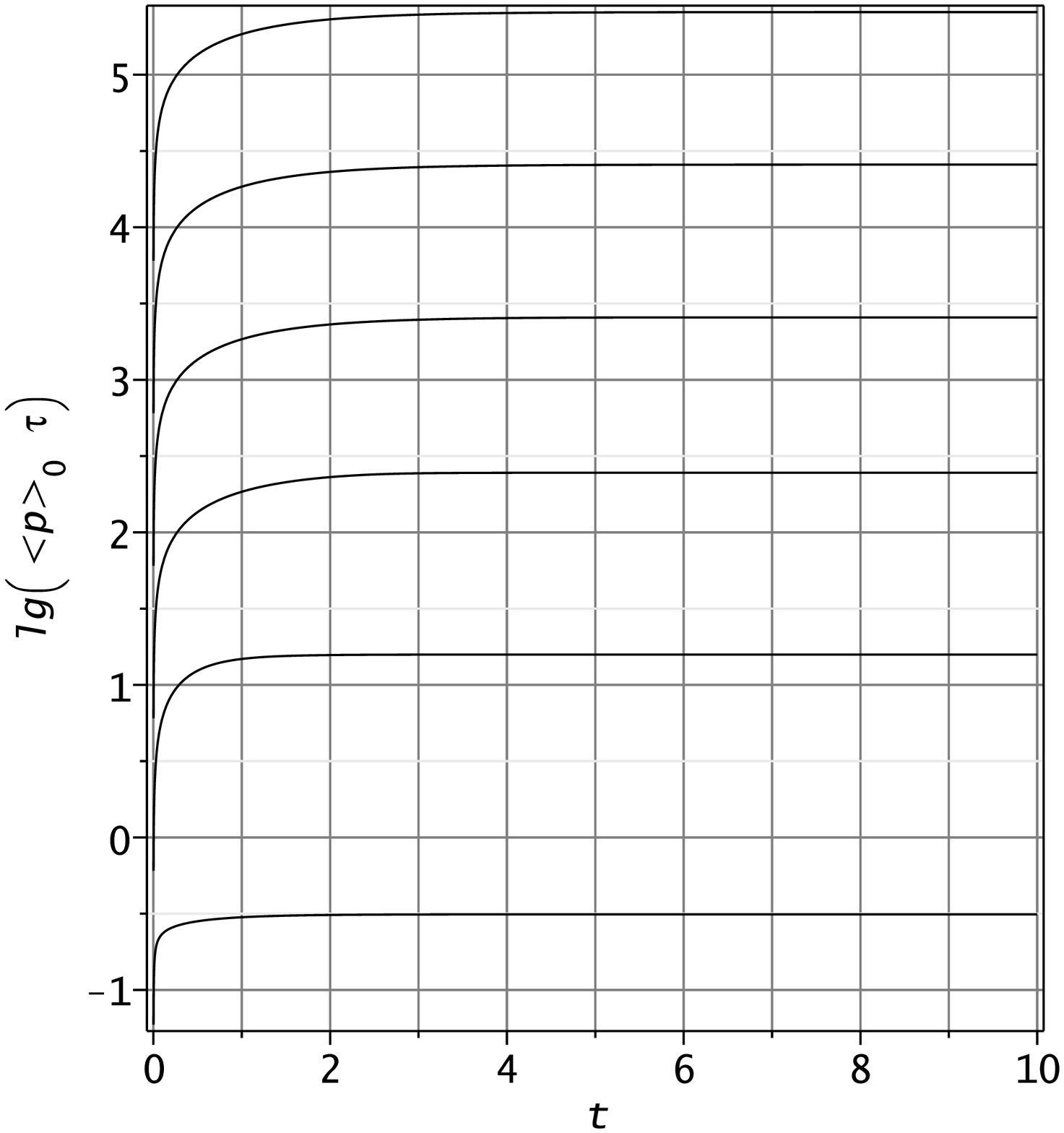}}
 \caption{Dependency of the dimensionless time variable's $\langle\tilde{p}\rangle_0\tau$ common logarithm
 on the cosmological time $t$. Bottom-up: $t_0=1$; $10$; $10^2$; $10^3$; $10^4$; $10^5$. It is introduced everywhere $N_0=100$; $N=10$.
 \label{tau_t_fig}}
\end{figure}

\begin{figure}[h]
 \centerline{\includegraphics[width=8cm]{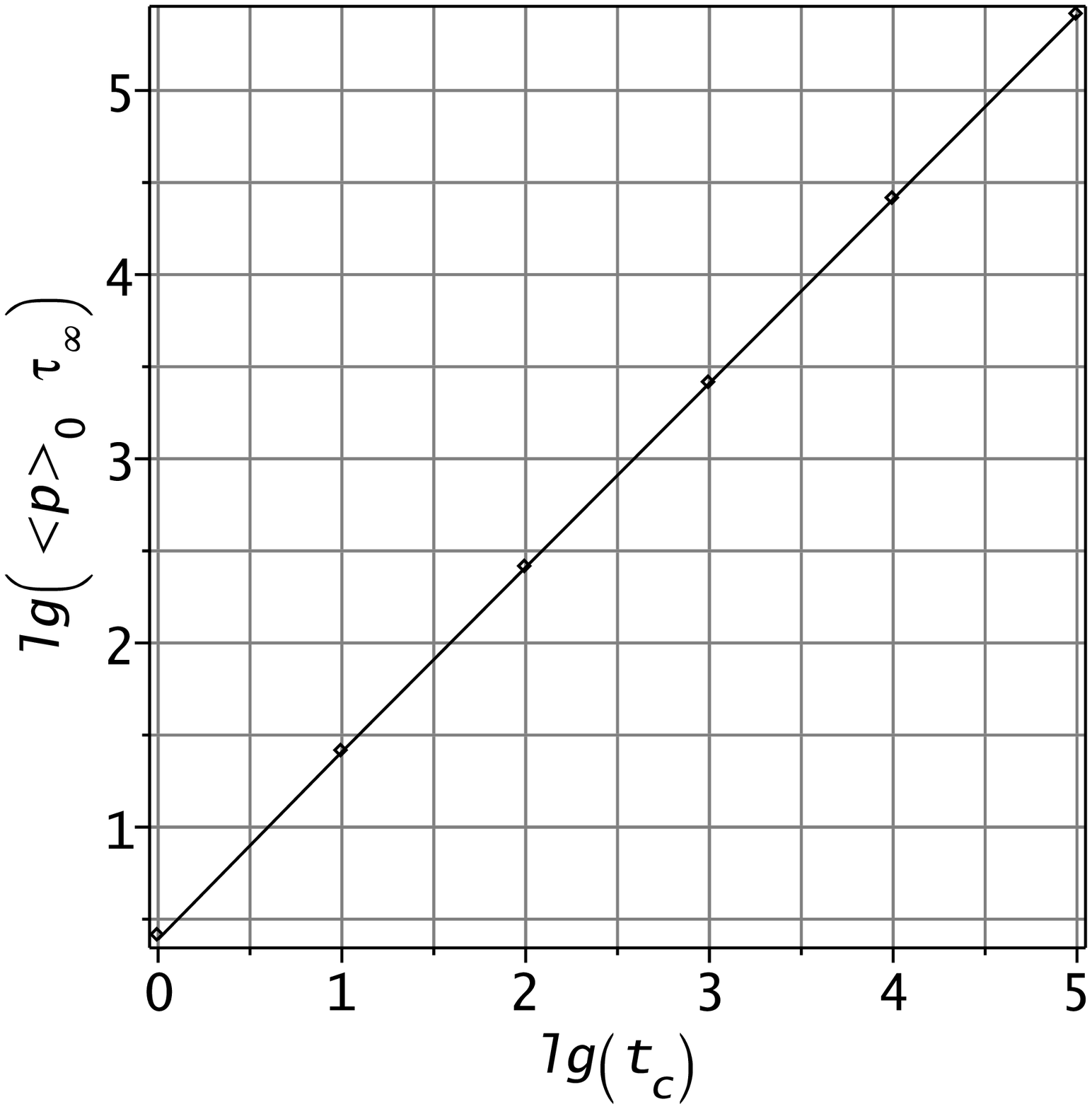}} \caption{Dependency of the dimensionless time parameter $\tau_\infty \langle\tilde{p}\rangle_0$
 on the time cosmological constant $t_0\equiv t_c$ (\ref{t_0}); everywhere is $t_c=1$; $N_0=100$; $N=10$, $\langle\tilde{p}\rangle_0$; $\sigma_0$ --- bottom-up:
 0.01; 0.1; 0.2; 0.5.
 \label{tau8_t0_fig}}
\end{figure}

\begin{figure}[h]
 \centerline{\includegraphics[width=8cm]{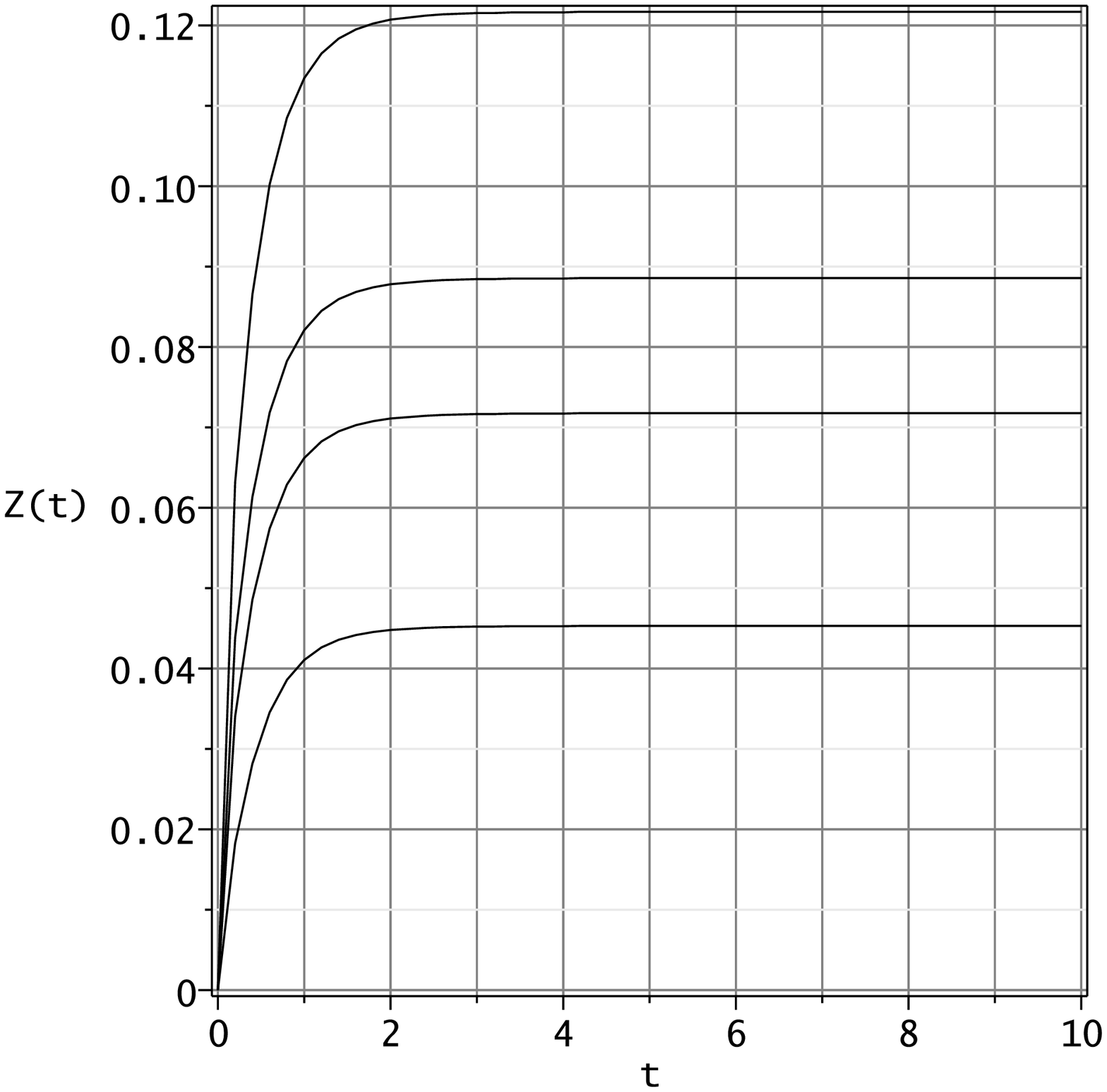}} \caption{Dependency of the dimensionless function $Z(t)$ on time; it is everywhere $t_c=1$; $N_0=100$; $N=10$, $\langle\tilde{p}\rangle_0=10$; $\sigma_0$ --- bottom up:
 0.01; 0.1; 0.2; 0.5.
 \label{Z_t_fig}}
\end{figure}

On Fig. \ref{tau_t_fig} the results of numerical integration for the
definition of the parameter $\tau_\infty$ are shown.
In particular, the integration of the relation (\ref{IV.42}) confirmed
insensitivity of  value $\tau_\infty$ from the number of parameters
and, practically, confirmed the esti\-ma\-tion formula (89) \cite{Ign1}
\begin{equation}\label{V.36}
\tau_\infty=\lim\limits_{t\to+\infty}\tau(t)=\frac{2\Lambda\langle\xi\rangle}{\langle\tilde{p}\rangle_0}
{\rm F}(1,1/\sqrt{2}),
\end{equation}
which did not account the details of the logarithmical dependency of
the para\-me\-ter $\langle\xi\rangle$ on time. On Fig. \ref{tau8_t0_fig}
the re\-sults of this value's numerical integration are shown.
These results are well described by formula
\begin{equation}\label{tau_inf}
\tau_\infty\approx \frac{2.57 t_0}{\langle\tilde{p}\rangle_0}.
\end{equation}
On Fig. \ref{Z_t_fig} the dependency of the variable Z(t) at
different values of time cosmological constant is shown.

\begin{figure}
 \centerline{\includegraphics[width=8cm]{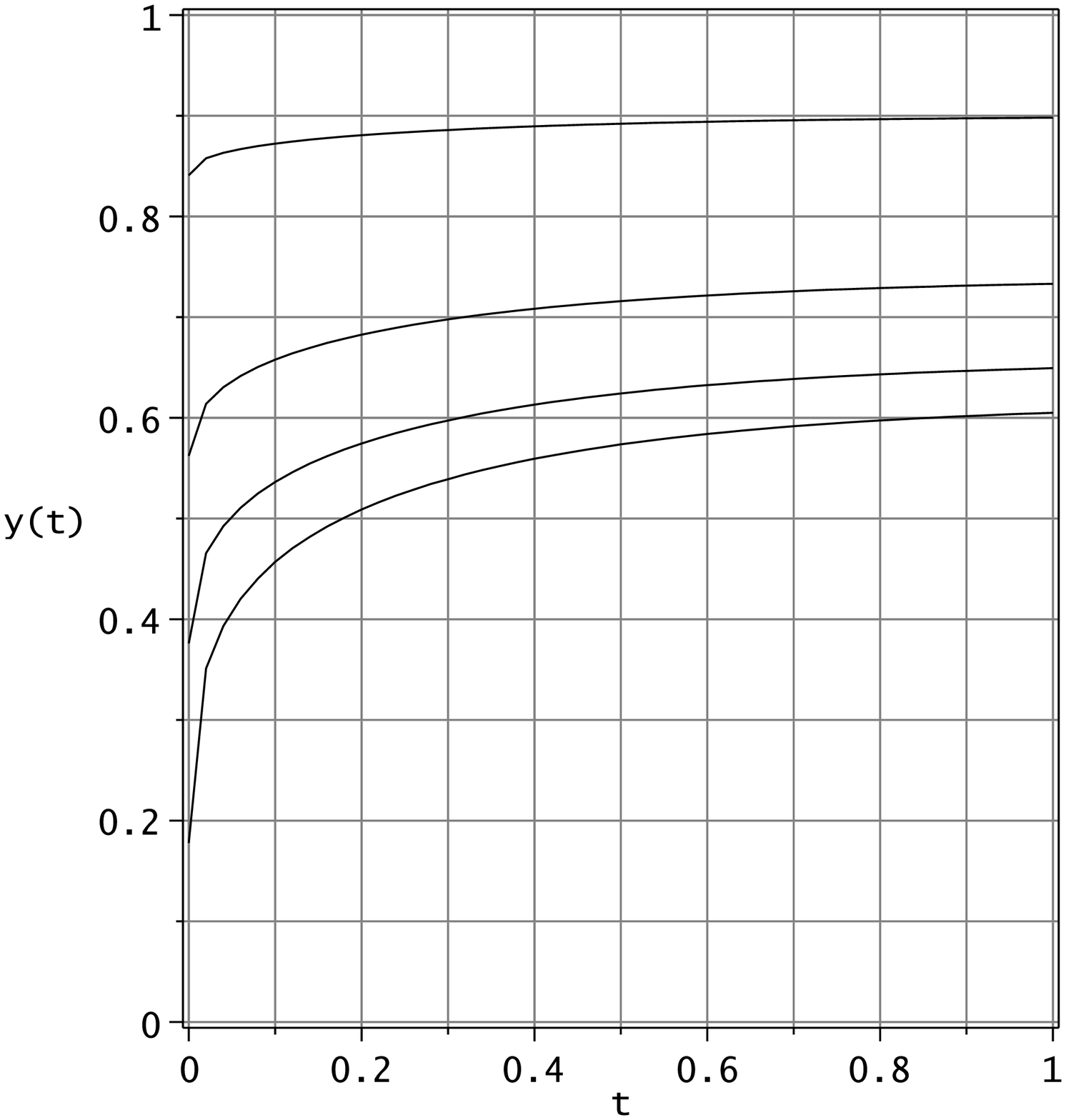}} \caption{Dependency of the relative temperature $y(t)$ on time; it is everywhere $t_c=1$; $N_0=100$; $N=10$, $\langle\tilde{p}\rangle_0=10$; $\sigma_0$ --- bottom up:
 0.01; 0.1; 0.2; 0.5.
 \label{y_t_fig}}
\end{figure}

As it is follows from the results represented on this figure,
function's  $Z(t)$ value also has a limit value at $t\to\infty$:
\begin{equation}\label{Z8}
\lim\limits_{t\to\infty}Z_\infty<\infty.
\end{equation}
According to (47) \cite{Ign1}
\begin{equation}\label{IV.45}
\Delta f_a(\tau,\rho)=\Delta f^0_a(\rho) \cdot{\rm
e}^{\displaystyle - \frac{Z(\tau)}{\rho}}.
\end{equation}
this means that at $t\to\infty$
superthermal particles' distribution is ``frozen'':
\begin{equation}\label{IV.45a}
\Delta f_a(\tau_\infty,\rho)=\Delta f^0_a(\rho) \cdot{\rm
e}^{\displaystyle - \frac{Z_\infty}{\rho}}.
\end{equation}
Thus, in modern Universe there can remain the ``tail'' of
non-equilibrium particles of extra-high energies:
\begin{equation}\label{E8}
E>E_\infty=Z_\infty\langle\tilde{p}\rangle_0.
\end{equation}

\begin{figure}[h]
 \centerline{\includegraphics[width=8cm]{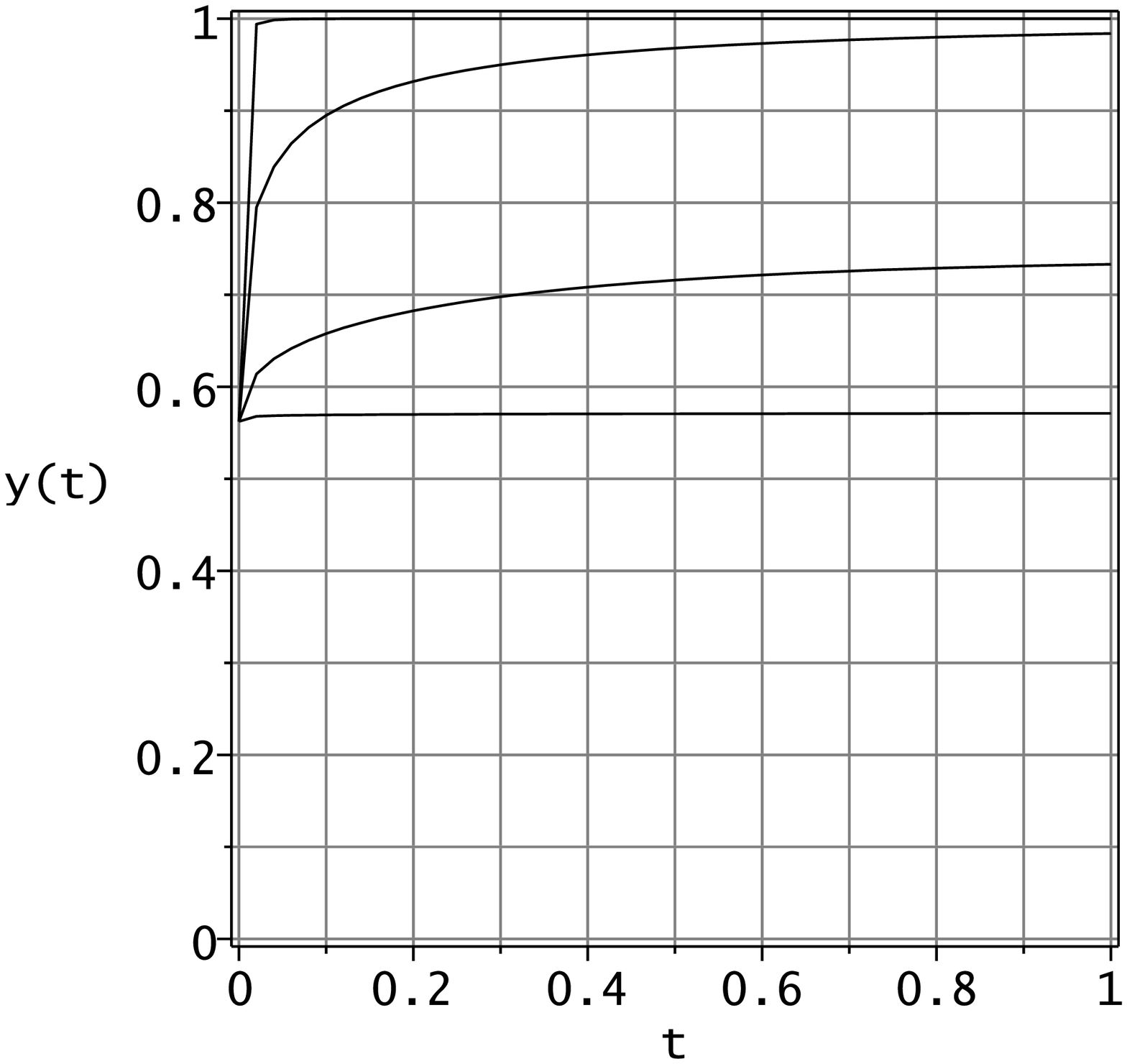}} \caption{Dependency of the relative temperature $y(t)$ on time; it is everywhere $\sigma_0=0.1$; $N_0=100$; $N=10$, $\langle\tilde{p}\rangle_0=10$;  $t_c$ --- bottom-up:
 0.1; 1; 10; 100.
 \label{y_s_fig}}
\end{figure}

\begin{figure}[h]
 \centerline{\includegraphics[width=8cm]{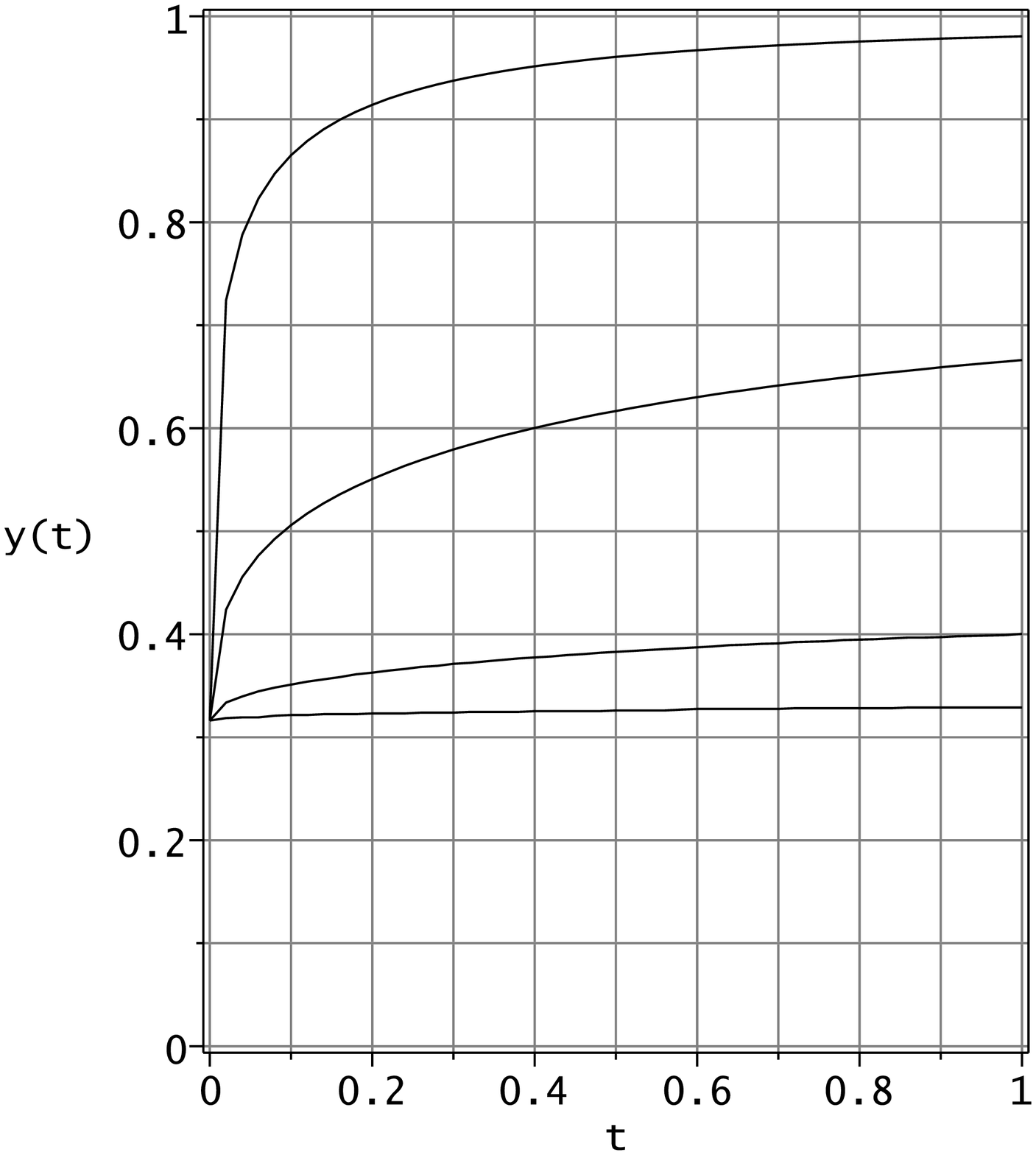}} \caption{Dependency of the relative temperature $y(t)$ on time; it is everywhere  $\sigma_0=0.01$; $N_0=100$; $N=10$, $\langle\tilde{p}\rangle_0=1000$; $t_0$ --- bottom-up:
 1; 10; 100; 1000.
 \label{y_t0_fig}}
\end{figure}

On Fig. \ref{y_s_fig}--\ref{y_t0_fig} the results of numerical
integration for the relative temperature $y(t)=T(t)/T_0(t)\leq1$ are
shown. According to the meaning of that value the dimensionless
parameter:
\begin{equation}\label{E_n8}
e_\infty=1-\sigma_\infty=1-y^4_\infty >0
\end{equation}
is a relative part of cosmological plasma's energy concluded in this
non-equilibrium ``tail'' of distri\-bu\-tion.

\subsection{Asymptotic values of parametres of nonequilibrium distribution at an inflationary stage}
\begin{figure}[h]
 \centerline{\includegraphics[width=8cm]{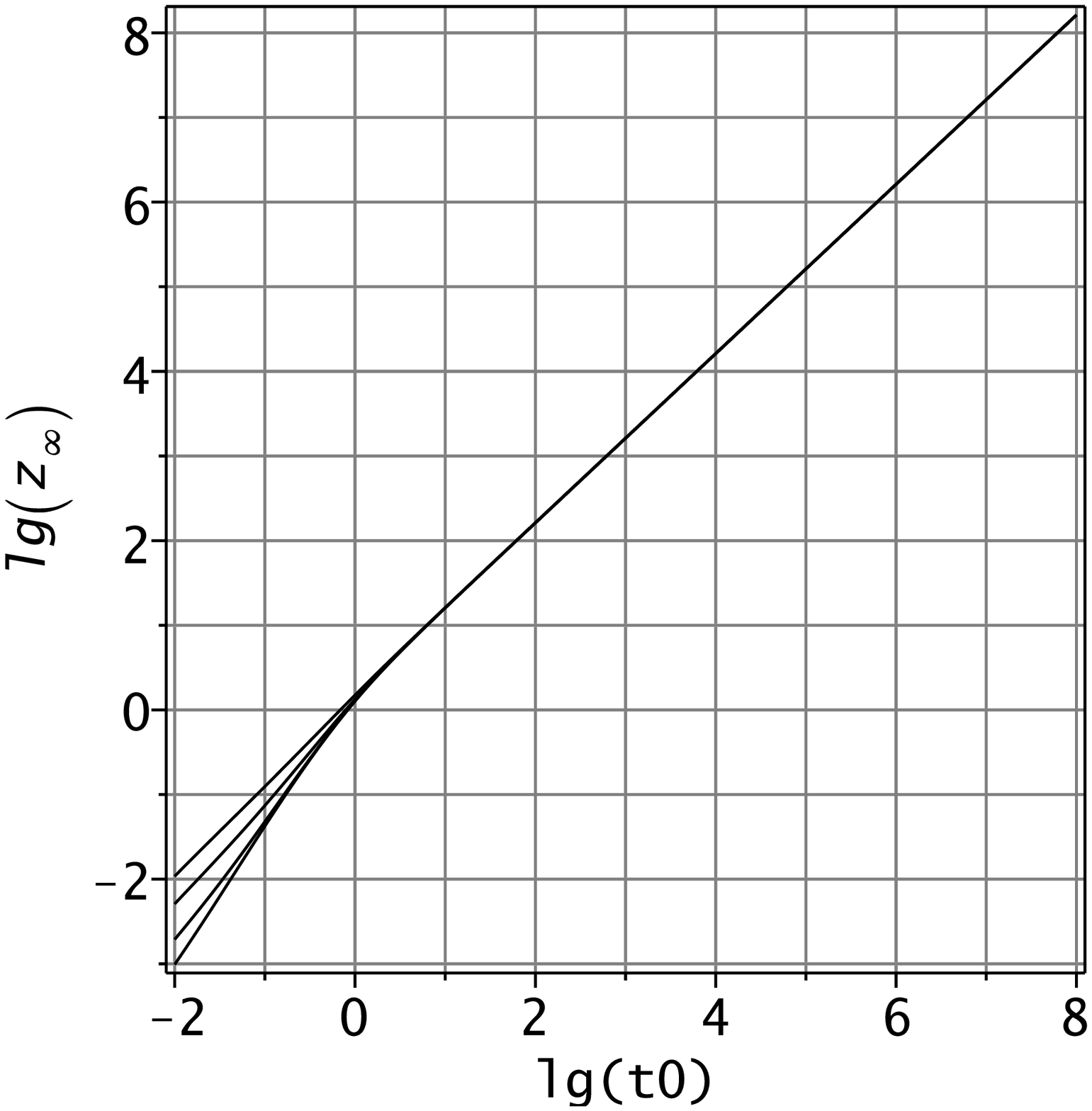}} \caption{Dependency of the dimensionless function $Z_\infty$ on cosmological constant $t_0$; it is everywhere $\langle\tilde{p}\rangle_0=100$; $N_0=100$; $N=10$;  $\sigma_0$ --- bottom-up:
 0.001; 0.01; 0.1; 0.5.
 \label{Z8_t0_fig}}
\end{figure}

\begin{figure}[h]
 \centerline{\includegraphics[width=8cm]{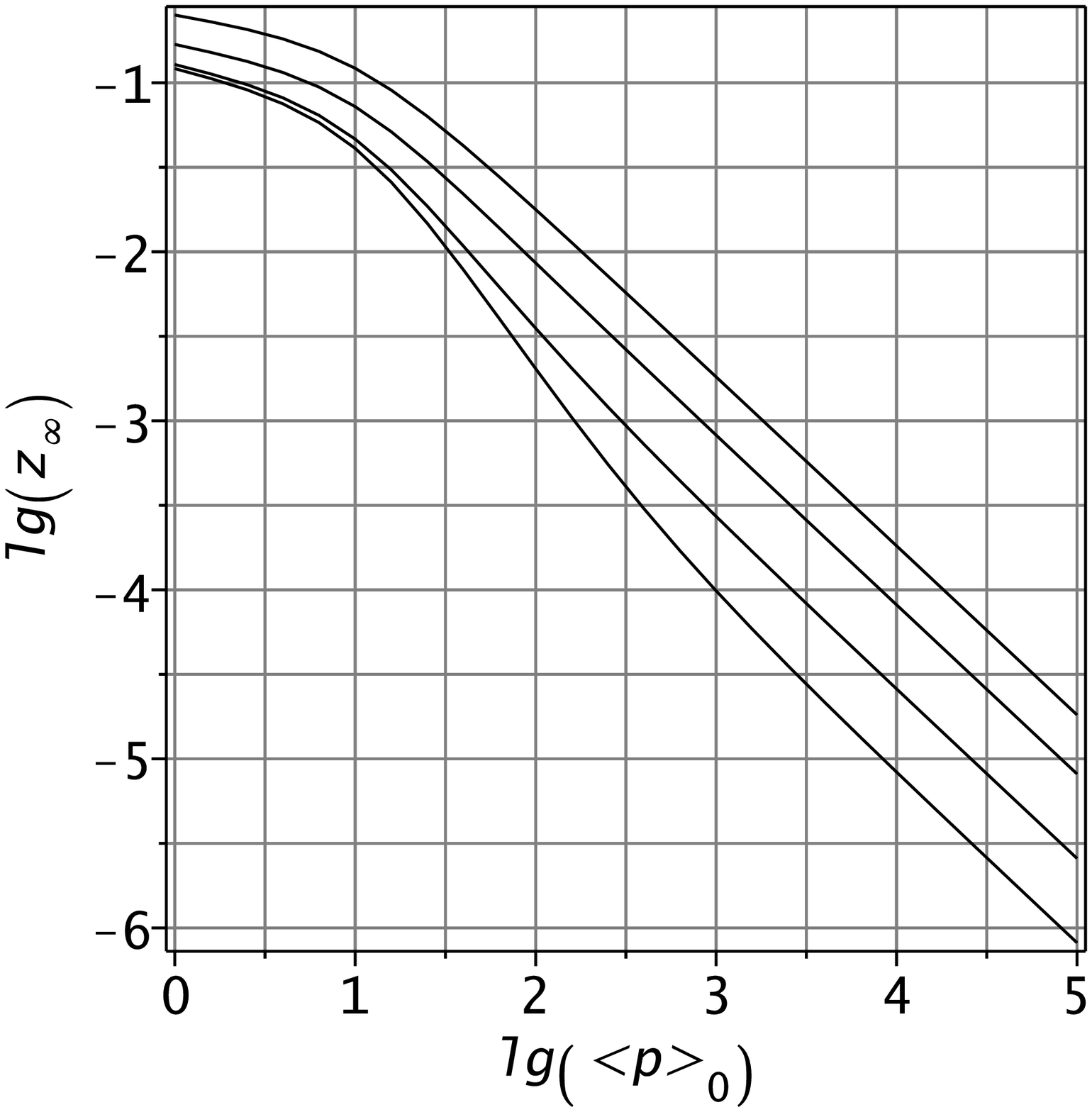}} \caption{Dependency of the dimensionless function $Z_\infty$ on parameter $\langle\tilde{p}\rangle_0$; it is everywhere $t_0=1$; $N_0=100$; $N=10$;  $\sigma_0$ --- bottom-up:
 0.001; 0.01; 0.1; 0.5.
 \label{Z8_lg_p0_fig}}
\end{figure}

\begin{figure}[h]
 \centerline{\includegraphics[width=8cm]{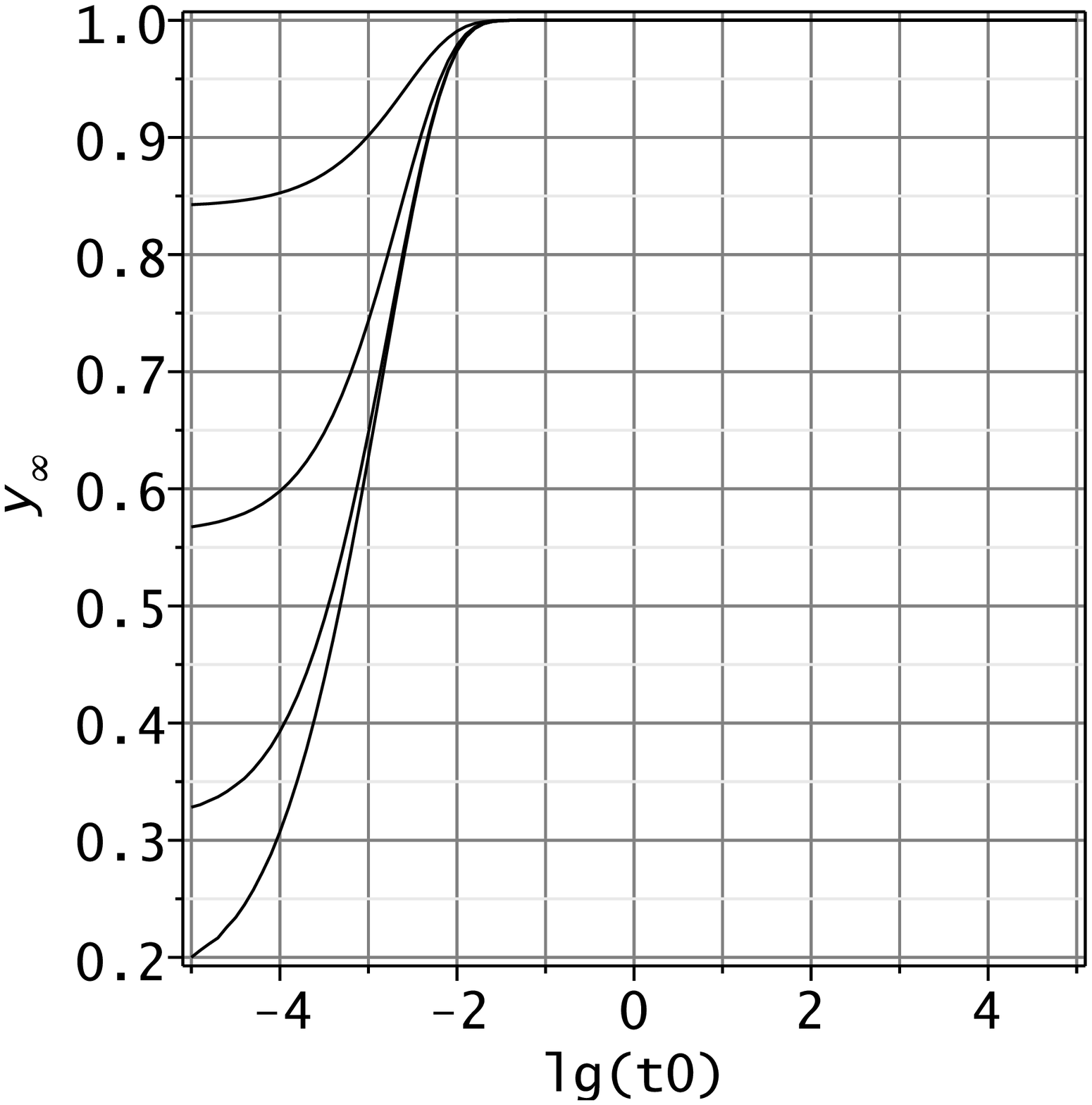}} \caption{Dependency of the relative temperature $y_\infty$ on cosmological constant $t_0$; it is everywhere $t_0=1$; $N_0=100$; $N=10$;  $\sigma_0$ --- bottom-up:
 0.001; 0.01; 0.1; 0.5.
 \label{y8_t0_fig}}
\end{figure}

\begin{figure}[h]
 \centerline{\includegraphics[width=8cm]{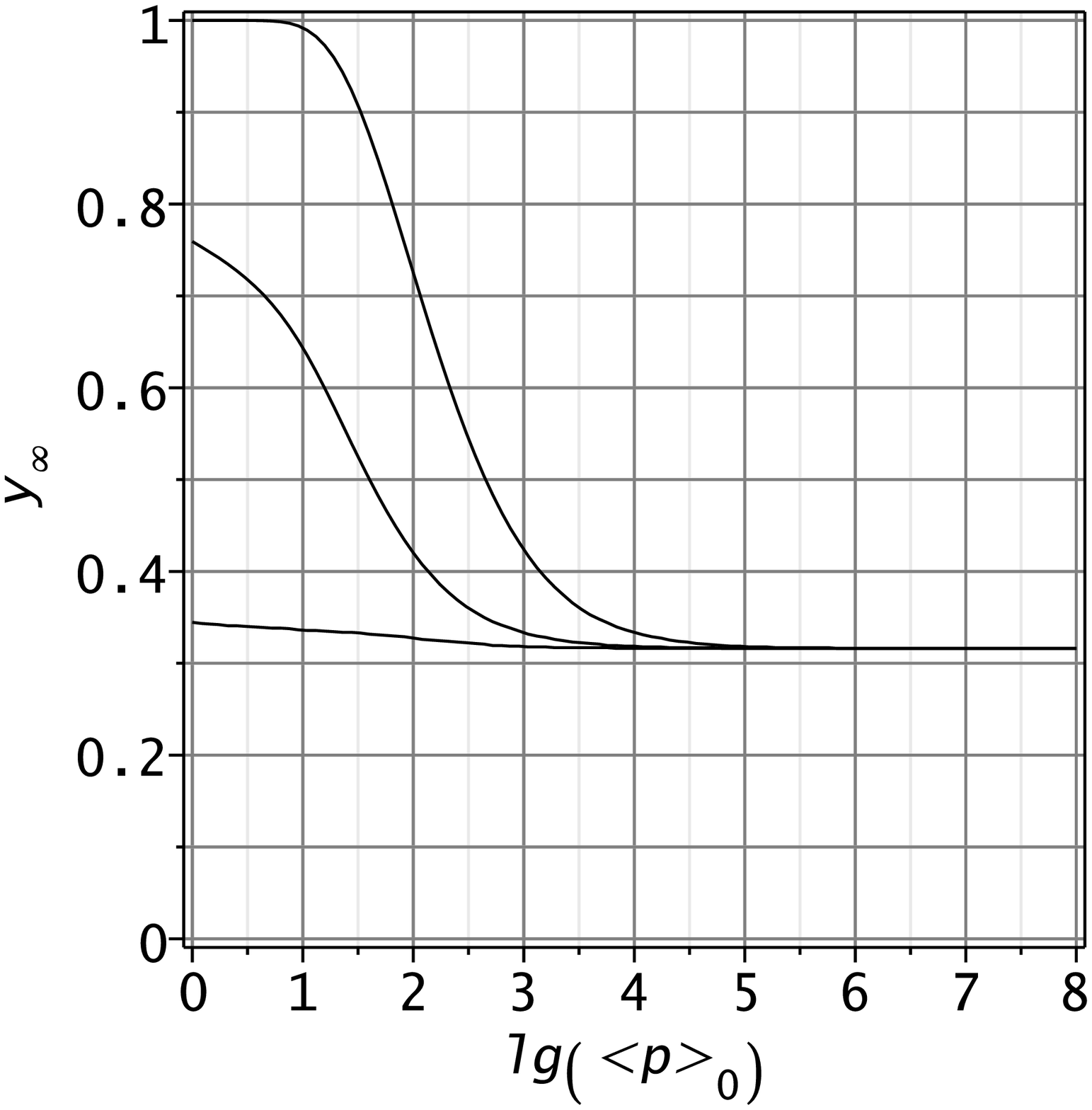}} \caption{of the relative temperature $y_\infty$ on parameter $\langle\tilde{p}\rangle_0$; it is everywhere $\sigma_0=0.1$; $N_0=100$; $N=10$;  $t_0$ --- bottom-up:
 0.1; 1; 10.
 \label{y8_p0_fig}}
\end{figure}

\begin{figure}[h]
 \centerline{\includegraphics[width=8cm]{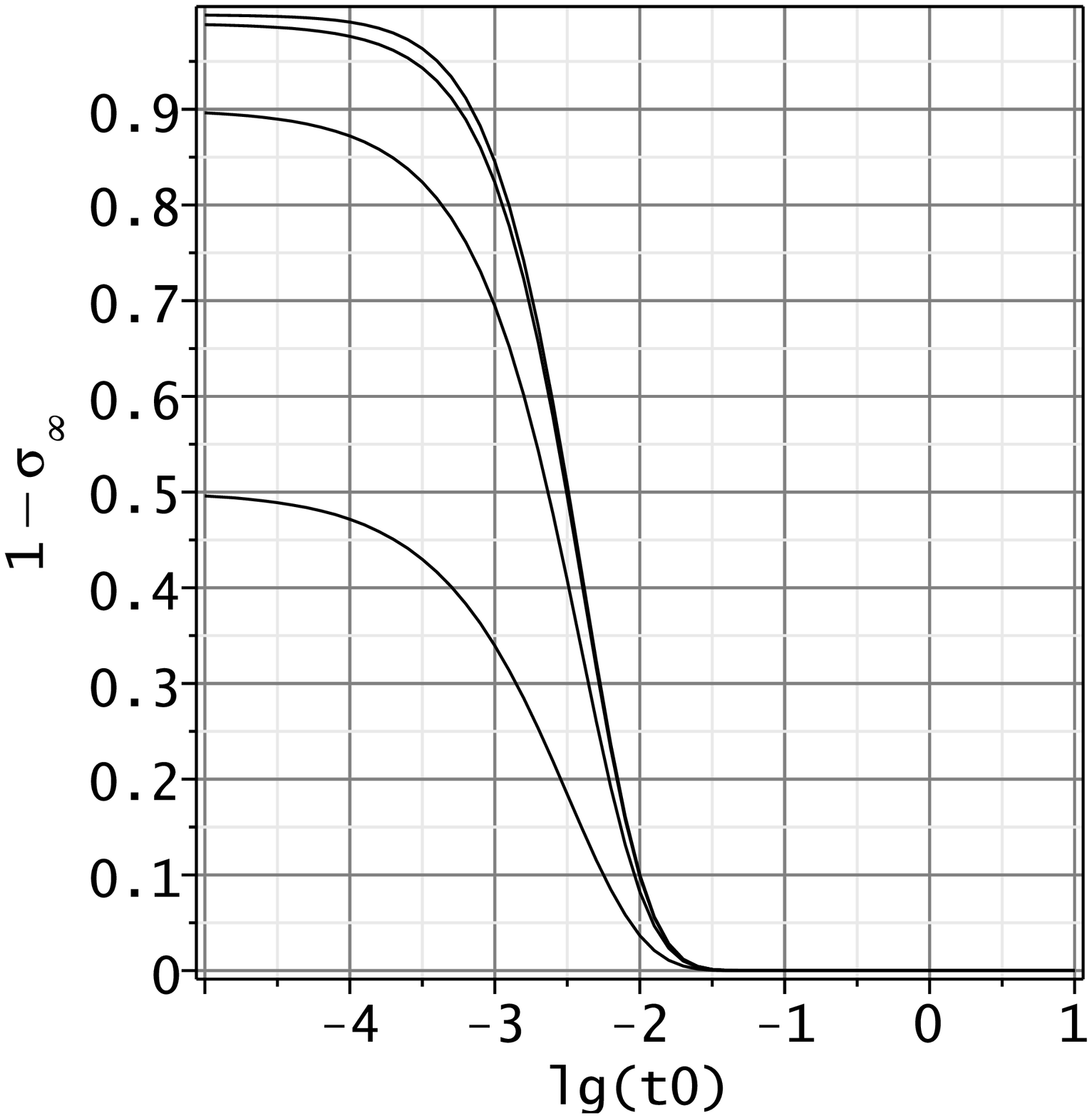}} \caption{Dependence of a relative part of energy of nonequilibrium particles, $1-\sigma_\infty$, on cosmological constant $t_0$; it is everywhere  $\langle\tilde{p}\rangle_0=1000$; $N_0=100$; $N=10$,  $\sigma_0$ --- top-down: 0.001; 0.01; 0.1; 0.5.
 \label{1-s8_t0_fig}}
\end{figure}

\begin{figure}[h]
 \centerline{\includegraphics[width=8cm]{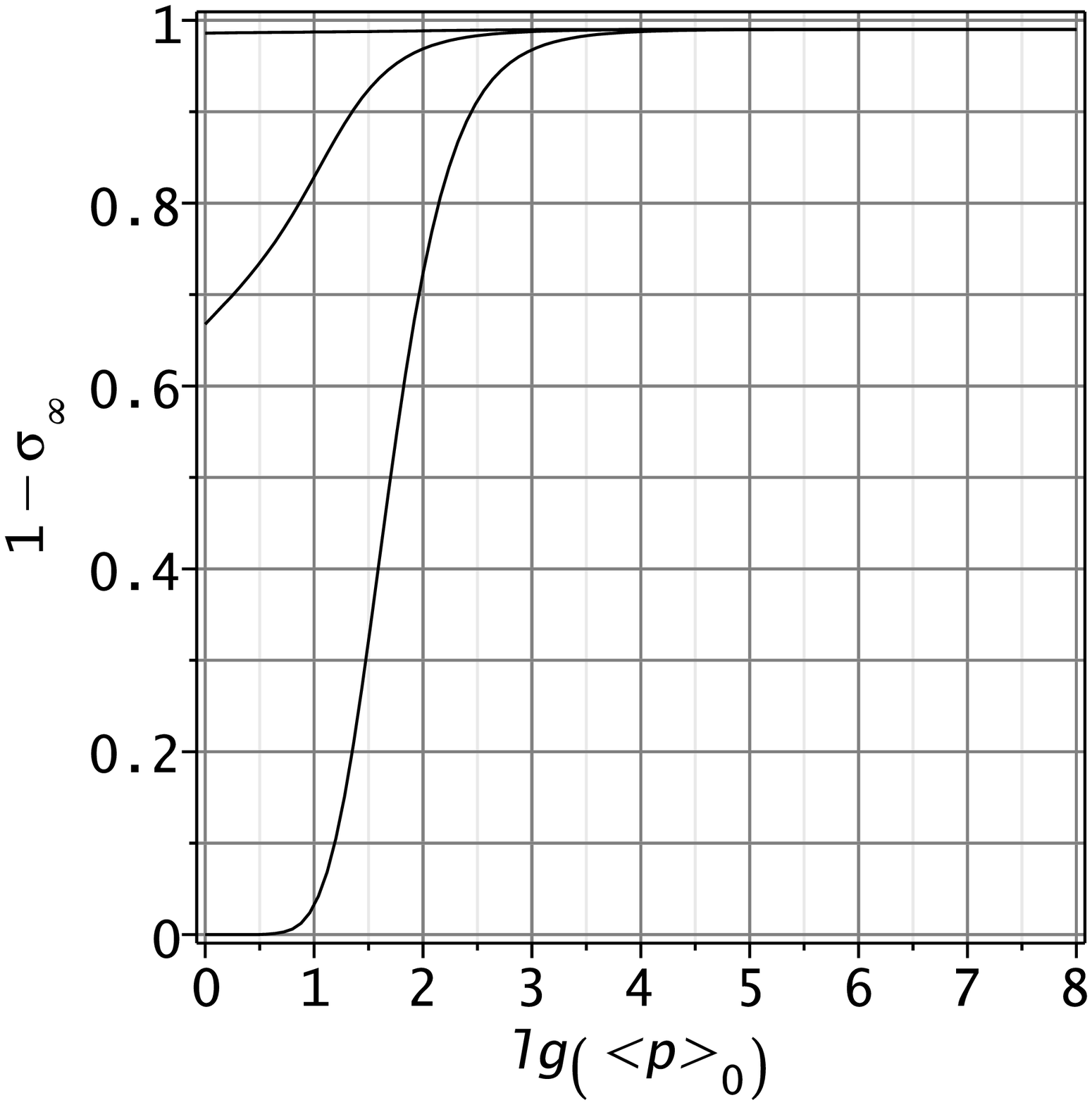}} \caption{Dependence of a relative part of energy of nonequilibrium particles, $1-\sigma_\infty$, on cosmological constant $t_0$; it is everywhere  $\sigma_0=0.01$; $N_0=100$; $N=10$, $\langle\tilde{p}\rangle_0=1000$; $t_0$ --- top-down: 0.1; 1; 10.
 \label{1-s8_p0_fig}}
\end{figure}

For supervision in present epoch the Universe the knowledge of possible limiting values of parametres of nonequilibrium distribution of particles is important. Such probable observable parametres are, first, relative temperature, $y_\infty$, the relative part of energy concluded in a nonequilibrium tail of distribution, and also the form of this distribution. On Fig.  \ref{Z8_t0_fig}, \ref{Z8_lg_p0_fig}, \ref{y8_t0_fig}, \ref{y8_p0_fig}, \ref{1-s8_t0_fig}, \ref{1-s8_p0_fig} the calculated values of the first two parametres are present.


\subsection{Analysis of numerical simulation}
From these results follows, that already at values of parameter $\langle\tilde{p}\rangle_0$ an order 10-100 the survival of considerable number of nonequilibrium relic particles at a modern stage of evolution of the Universe is probable. It is the striking fact if to recollect, that according to results of early works of the Author in the standard cosmological scenario, deprived of an inflationary stage, at a modern stage only relic particles with energy of an order $10^12$Gev and above can survive. At presence of a modern inflationary stage nonequilibrium relic particles with energy of an order 1 ev can survive!

\begin{figure}[h]
 \centerline{\includegraphics[width=8cm]{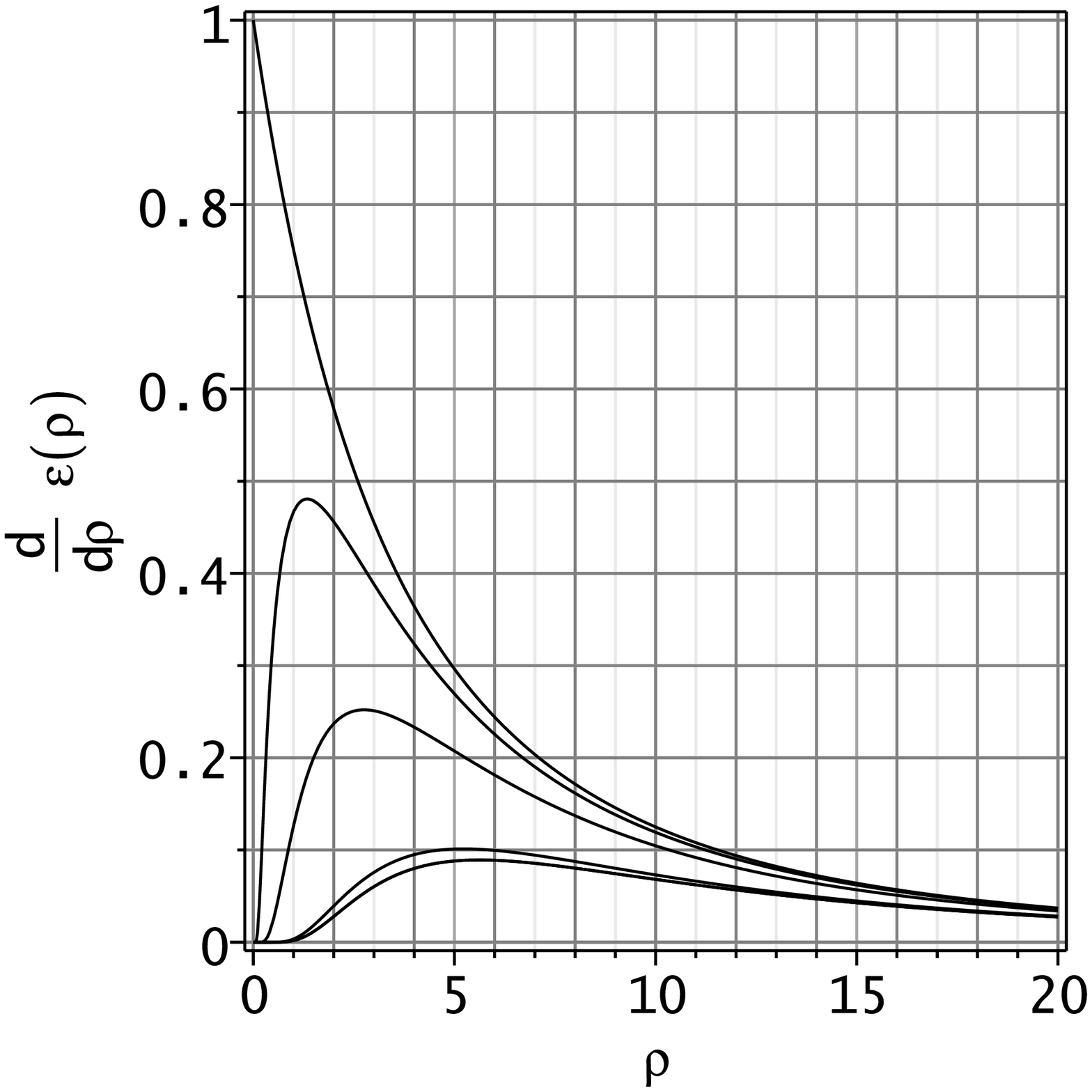}} \caption{Dependency of the relative temperature $y(t)$ on time; it is everywhere  $\sigma_0=0.01$; $N_0=100$; $N=10$, $\langle\tilde{p}\rangle_0=1000$; $t_0$ --- bottom-up:
 1; 10; 100; 1000.
 \label{de_dr_t_0001_100}}
\end{figure}

On Fig. \ref{de_dr_t_0001_100} the evolution of distribution of density of energy of nonequilibrium particles in the assumption of their initial distribution in form
\begin{equation}\label{de0_1}
\left(\frac{d\varepsilon}{d\rho}\right)_0= A\frac{\chi(\rho_0-\rho)}{(1+\rho/\langle\tilde{p}\rangle_0)^3}
\end{equation}
is present.

\begin{figure}[h]
 \centerline{\includegraphics[width=8cm]{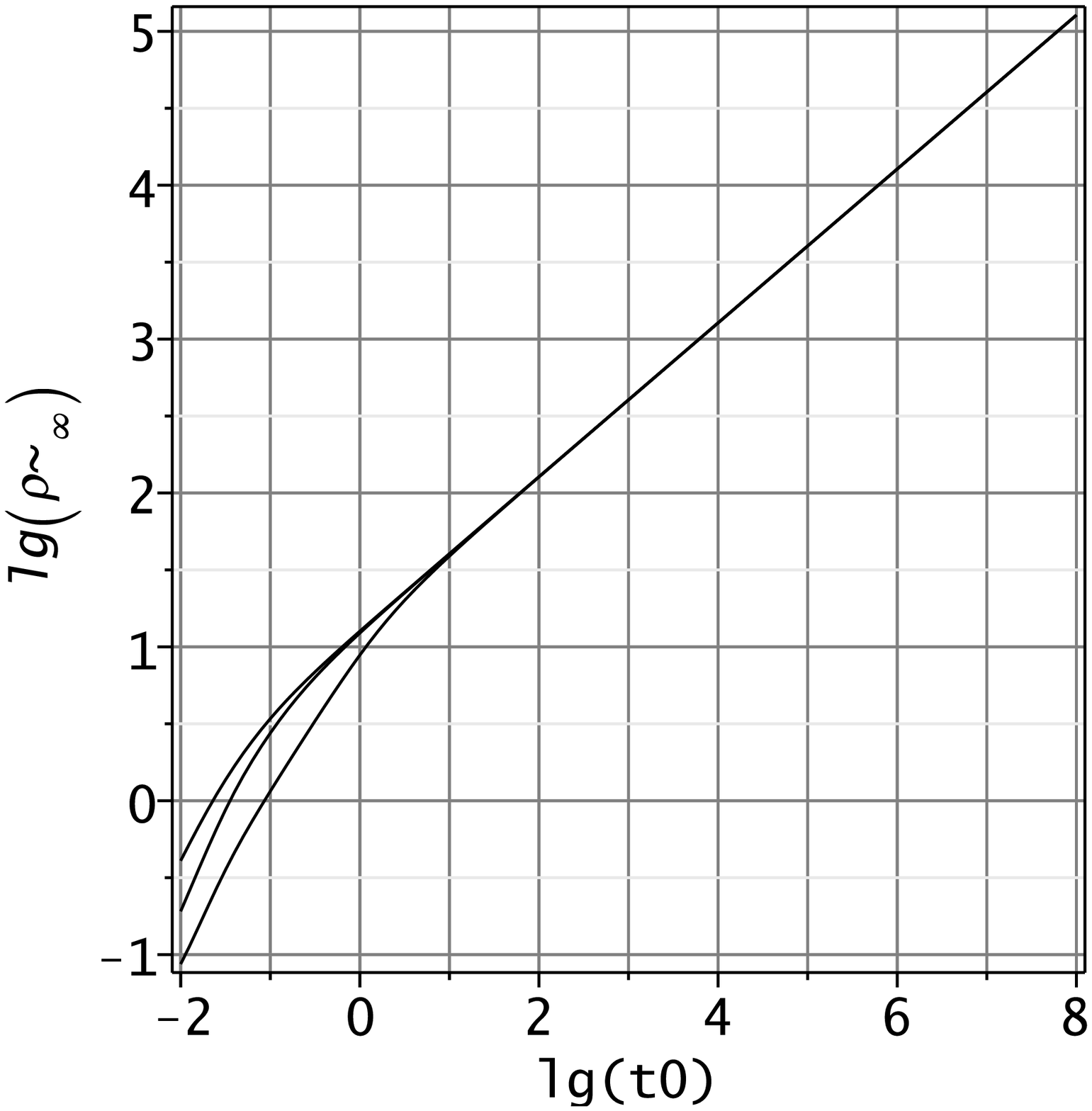}} \caption{Dependency of the relative temperature $y(t)$ on time; it is everywhere  $\sigma_0=0.01$; $N_0=100$; $N=10$, $\langle\tilde{p}\rangle_0=1000$; $t_0$ --- bottom-up:
 1; 10; 100; 1000.
 \label{rho8_t0}}
\end{figure}

Calculating a maximum of distribution of density of energy from a relationship (\ref{IV.45}), we will find:
\begin{equation}\label{e_max}
\rho_\infty^{max}=\sqrt{Z_\infty \langle\tilde{p}\rangle_0}.
\end{equation}
On Fig. \ref{rho8_t0} dependence of a maximum of a energy spectrum of nonequilibrium particles on cosmological constant $t_0$ is present. Apparently from this figure, at not so great values of a cosmological constant the maximum of a energy spectrum of nonequilibrium particles lays in the field of enough low energies, that, of course, does probable their detection in space conditions.

\section*{Conclusion}

In the conclusion the Author once again expresses gratitude to professor Vitaly
Melnikov for extremely fruitful question.

\end{document}